\documentclass[preprint]{JHEP3} 

\JHEPspecialurl{http://jhep.sissa.it/JOURNAL/JHEP3.tar.gz}

\usepackage{epsfig,amsmath,enumerate}

\newcommand\fverb{\setbox\fverbbox=\hbox\bgroup\verb}
\newcommand\fverbdo{\egroup\medskip\noindent%
            \fbox{\unhbox\fverbbox}\ }
\newcommand\fverbit{\egroup\item[\fbox{\unhbox\fverbbox}]}
\newbox\fverbbox

\title{Shear Dynamics in Bianchi I Cosmology}

\author{Diego L. C\'{a}ceres, Leonardo Casta\~{n}eda and Juan M. Tejeiro \\
    Observatorio Astron\'omico Nacional, Universidad Nacional de Colombia, Bogota, Colombia\\
    E-mail: \email{dlcaceresu@bt.unal.edu.co}, \email{lcastanedac@unal.edu.co} and
    \email{jmtejeiros@unal.edu.co}

}

\abstract{We present the exact equation for evolution of Bianchi I
cosmological model, considering a non-tilted perfect fluid in a matter dominated universe. We use
the definition of shear tensor and later we prove it is consistent
with the evolution equation for shear tensor obtained from Ricci
identities and widely known in literature \cite{9812046},
\cite{0705.4397v3}, \cite{wainwright}. Our result is compared with
the equation given by Ellis and van Elst in \cite{9812046} and
Tsagas, Challinor and Maartens \cite{0705.4397v3}. We consider
that it is important to clarify the notation used in \cite{9812046}, \cite{0705.4397v3} related with the covariant derivative and the behavior of the shear tensor.}

\begin{document}


\section{1+3 Orthonormal frame approach}

In a cosmological space-time $(\mathcal{M},\mathbf{g})$ there are
preferred worldlines representing the average motion of matter at
each point, associated with comoving or fundamental observers,
which do not have peculiar velocities. The signature used in this article is $(-+++)$. The 4-velocity of the comoving particles is $u^{\alpha}$, $u^{\alpha}=(1,0,0,0)$, $u^{\alpha}u_{\alpha}=-1$. This
4-velocity is orthogonal to the surfaces of spatial homogeneity.
Therefore, it is defined the spatial projection tensor
$h_{\alpha\beta}$ as \cite{9812046}, \cite{ehlers}:

\begin{equation}
h_{\alpha\beta}:=g_{\alpha\beta}+u_{\alpha}u_{\beta},
\end{equation}

\noindent given this tensor, we can define the orthogonally projected
symmetric trace-free part of any tensor $T^{\alpha\beta}$ of
second rank as:

\begin{equation}
T^{\langle\alpha\beta\rangle}=\left[h^{(\alpha}_{\ \
\gamma}h^{\beta)}_{\ \
\delta}-\frac{1}{3}h^{\alpha\beta}h_{\gamma\delta}\right]T^{\gamma\delta},
\end{equation}

\noindent moreover, other two derivatives can be defined \cite{9812046}: the
\textit{covariant time derivative}, along the fundamental
worldlines, where for any tensor $T^{\alpha\beta}_{\ \
\gamma\delta}$:

\begin{equation}\label{covariant time derivative}
\dot{T}^{\alpha\beta}_{\ \
\gamma\delta}:=u^{\epsilon}\nabla_{\epsilon}T^{\alpha\beta}_{\ \
\gamma\delta},
\end{equation}

\noindent and the fully \textit{orthogonally projected covariant derivative}
$\tilde{\nabla}$, where:

\begin{equation}
\tilde{\nabla}_{\epsilon}T^{\alpha\beta}_{\ \ \gamma\delta}=h^{\alpha}_{\
\mu}h^{\beta}_{\ \nu}h^{\kappa}_{\ \gamma}h^{\lambda}_{\
\delta}h^{\eta}_{\ \epsilon}\nabla_{\eta}T^{\mu\nu}_{\ \ \kappa\lambda}.
\end{equation}

\noindent Given these derivatives, the 4-aceleration can be written as $\dot{u}_{\alpha}=u_{\alpha;\beta}u^{\beta}$. With these
definitions the first covariant derivative of
$u_{\alpha}$ is decomposed into its irreducible parts, defined by their symmetry
properties \cite{9812046}, \cite{ehlers}:

\begin{equation}
\nabla_{\beta}u_{\alpha}=-\dot{u}_{\alpha}u_{\beta}+\omega_{\alpha\beta}+\sigma_{\alpha\beta}+\frac{1}{3}\Theta
h_{\alpha\beta},
\end{equation}

\noindent where $\sigma_{\alpha\beta}:=u_{(\alpha;\beta)}-\dot{u}_{(\alpha}u_{\beta)}-\frac{1}{3}\Theta
h_{\alpha\beta}$ is the trace-free symmetric \textit{rate of
shear} tensor $(\sigma_{\alpha\beta}=\sigma_{(\alpha\beta)},\
\sigma_{\alpha\beta}u^{\beta}=0,\ \sigma^{\alpha}_{\ \alpha}=0)$,
which describes the rate of distortion of the matter flow; and
$\omega_{\alpha\beta}:=u_{[\alpha;\beta]}-\dot{u}_{[\alpha}u_{\beta]}$ is
the skew-symmetric \textit{vorticity} tensor
$(\omega_{\alpha\beta}=\omega_{[\alpha\beta]},\
\omega_{\alpha\beta}u^{\beta}=0)$, describing the rotation of the
matter relative to a non-rotating (Fermi-propagated) frame \cite{9812046}.\\

\noindent It is possible to obtain a propagation equation for the shear
tensor, from Ricci identities \cite{9812046}, \cite{0705.4397v3}:

\begin{equation}\label{shearevolution}
\dot{\sigma}^{\langle\alpha\beta\rangle}-u^{\langle\alpha}u^{\beta\rangle}=-\frac{2}{3}\Theta\sigma^{\alpha\beta}+\dot{u}^{\langle\alpha}\dot{u}^{\beta\rangle}-\sigma^{\langle\alpha}_{\gamma}\sigma^{\beta\rangle\gamma}-\omega^{\langle\alpha}\omega^{\beta\rangle}
-\left(E^{\alpha\beta}-\frac{1}{2}\pi^{\alpha\beta}\right),
\end{equation}

\noindent where $E^{\alpha\beta}$ is the Electric Weyl tensor,
$E_{\alpha\beta}=C_{\alpha\gamma\beta\delta}u^{\gamma}u^{\delta}$,
where $C_{\alpha\beta\gamma\delta}$ is the Weyl tensor, and
$\pi_{\alpha\beta}=T_{\gamma\delta}h^{\gamma}_{\
\langle\alpha}h^{\delta}_{\ \beta\rangle}$, where
$T_{\alpha\beta}$ is the \textit{energy momentum tensor} and $\pi_{\alpha\beta}$ is the trace-free \textit{anisotropic pressure}.\\

\noindent The Weyl Tensor is completely determined from its electric and magnetic parts, the last one defined as \cite{9812046}, \cite{van Elst thesis} defined as:

\begin{equation}
H_{\alpha\beta}=\frac{1}{2}\eta_{\alpha\delta\epsilon}C^{\delta\epsilon}_{\ \ \beta\gamma}u^{\gamma},
\end{equation}

\noindent where $\eta_{\alpha\beta\gamma}=u^{\delta}\eta_{\delta\alpha\beta\gamma}$ is a \textit{volume element} for the rest spaces and $\eta_{\alpha\beta\gamma\delta}$ is the 4-dimensional volume element \cite{9812046} $(\eta_{\alpha\beta\gamma\delta}=\eta_{[\alpha\beta\gamma\delta]},\ \eta_{0123}=\sqrt{|\det(g_{\alpha\beta})|})$.

\noindent Using the Gauss-Codacci relation the Spatial Riemann Tensor is \cite{0705.4397v3}:

\begin{equation}
^{3}R_{\alpha\beta\gamma\delta}=h^{\zeta}_{\ \alpha}h^{\eta}_{\
\beta}h^{\theta}_{\ \gamma}h^{\vartheta}_{\
\delta}R_{\zeta\eta\theta\vartheta}-\tilde{\nabla}_{\gamma}u_{\alpha}\tilde{\nabla}_{\delta}u_{\beta}+\tilde{\nabla}_{\delta}u_{\alpha}\tilde{\nabla}_{\gamma}u_{\beta}.
\end{equation}

\noindent and the Spatial Ricci tensor is:

\begin{equation}\label{Ricci}
^{3}R_{\alpha\beta}=-\dot{\sigma}_{\langle
\alpha\beta\rangle}-\Theta\sigma_{\alpha\beta}+\tilde{\nabla}_{\langle\alpha}\dot{u}_{\beta\rangle}+\dot{u}_{\langle\alpha}\dot{u}_{\beta\rangle}+
\pi_{\alpha\beta}+\frac{1}{3}h_{\alpha\beta}\left[2\mu-\frac{2}{3}\Theta^{2}+2\sigma^{2}+2\Lambda\right],
\end{equation}

\noindent where $\mu$ is the energy density and
$\sigma^{2}=\frac{1}{2}\sigma^{\alpha\beta}\sigma_{\alpha\beta}$.

\section{Bianchi I cosmology}

Bianchi cosmologies are spatially homogeneous but not necessarily
isotropic. For a review of Bianchi models, see
\cite{CommmathPhys12_108-41}, \cite{CommmathPhys19_31-64},
\cite{wainwright} and for orthonormal frame approach \cite{9812046}, \cite{van
Elst thesis}, \cite{wainwright}.\\

\noindent Here we will consider Bianchi I cosmology. The metric of this model is
given by \cite{van Elst thesis}, \cite{0705.4397v3},
\cite{campanelli}, \cite{cea}:\\

\begin{equation}
ds^{2}=-dt^{2}+a^{2}(t)dx^{2}+b^{2}(t)dy^{2}+c^{2}(t)dz^{2},
\end{equation}

\noindent and the \textit{average expansion scale factor}
$S(t)=\sqrt[3]{abc}$. It reduces to the FLRW case when
$a(t)=b(t)=c(t)=S(t)$. Given this metric the
connection components are:

\begin{equation}\label{connection}
\begin{split}
&\Gamma^{t}_{xx}=a\dot{a},\qquad\Gamma^{x}_{xt}=\frac{\dot{a}}{a},\\
&\Gamma^{t}_{yy}=b\dot{b},\qquad\ \Gamma^{y}_{yt}=\frac{\dot{b}}{b},\\
&\Gamma^{t}_{zz}=c\dot{c},\qquad\ \Gamma^{z}_{zt}=\frac{\dot{c}}{c}.
\end{split}
\end{equation}

\noindent We are going to study the dynamic evolution of shear tensor from these connection components.

\subsection{Shear Dynamics}

The solution for the scale factors can be obtained directly from
Einstein Equations when we consider a perfect fluid
\cite{9812046}:

\begin{equation}\label{a}
a(t)=S(t)\exp(\Sigma_{1}W(t)),
\end{equation}

\begin{equation}\label{b}
b(t)=S(t)\exp(\Sigma_{2})W(t),
\end{equation}

\begin{equation}\label{c}
c(t)=S(t)\exp(\Sigma_{3}W(t)),
\end{equation}

\noindent where

\begin{equation}
W(t)=\int\frac{dt}{S^{3}(t)},
\end{equation}

\noindent and the constants $\Sigma_{\alpha}$ satisfy \cite{9812046}:

\begin{equation}
\Sigma_{1}+\Sigma_{2}+\Sigma_{3}=0.
\end{equation}

\noindent Now, using the shear tensor definition we get:

\begin{equation}
\sigma_{xx}=\frac{2}{3}a\dot{a}-\frac{a^{2}}{3}\frac{\dot{b}}{b}-\frac{a^{2}}{3}\frac{\dot{c}}{c},
\end{equation}

\begin{equation}
\sigma_{yy}=\frac{2}{3}b\dot{b}-\frac{b^{2}}{3}\frac{\dot{a}}{a}-\frac{b^{2}}{3}\frac{\dot{c}}{c},
\end{equation}

\begin{equation}
\sigma_{zz}=\frac{2}{3}c\dot{c}-\frac{c^{2}}{3}\frac{\dot{a}}{a}-\frac{c^{2}}{3}\frac{\dot{b}}{b}.
\end{equation}

\noindent From \eqref{a}, \eqref{b} and \eqref{c} we obtain:

\begin{equation}\label{shearx}
\sigma_{xx}=\Sigma_{1}\frac{a^{2}(t)}{S^{3}(t)},
\end{equation}

\begin{equation}\label{sheary}
\sigma_{yy}=\Sigma_{2}\frac{b^{2}(t)}{S^{3}(t)},
\end{equation}

\begin{equation}\label{shearz}
\sigma_{zz}=\Sigma_{3}\frac{c^{2}(t)}{S^{3}(t)}.
\end{equation}

\noindent Now, the spatial homogeneity of the Bianchi I space-times ensures
that all invariants depend at most on time. It is an irrotational
universe, $\omega_{\alpha\beta}=0$, and also it is spatially flat,
$^{3}R_{\alpha\beta}=0$. With a perfect fluid, $\pi_{\alpha\beta}=0$,
the equation \eqref{Ricci} can be reduced to \cite{9812046},
\cite{0705.4397v3}:

\begin{equation}\label{sheardynamics}
\dot{\sigma}_{\alpha\beta}=-3\frac{\dot{S}}{S}\sigma_{\alpha\beta}
\end{equation}

\noindent Following this equation, it seems that Van Elst and Ellis in \cite{9812046} and
Tsagas, Challinor and Maartens in \cite{0705.4397v3} conclude that
in the absence of anisotropic pressures the shear behaves as
$S^{-3}$. From \eqref{sheardynamics} it is easy to give a
non-correct interpretation for the shear dynamics because we could
conclude shear tensor is equal to a constant times $S^{-3}$, but
from \eqref{shearx}, \eqref{sheary} and \eqref{shearz} we see the
scale factors play a role in shear dynamics.\\
For checking our result we verify our shear expression \eqref{shearx}, \eqref{sheary} and \eqref{shearz} is
consistent with shear evolution equation \eqref{shearevolution}.
Given the definition of covariant derivative we have:

\begin{equation}
\dot{\sigma}_{xx}=\frac{\partial\sigma_{xx}}{\partial
t}-2\frac{\dot{a}}{a}\sigma_{xx}.
\end{equation}

\noindent It is not immediate to integrate this equation to get $\sigma\
\alpha\ S^{-3}(t)$. Taking into account this term in the evolution
equation:

\noindent
\begin{equation}
\frac{\partial\sigma_{xx}}{\partial
t}=\left(\frac{\dot{a}}{a}-\frac{\dot{b}}{b}-\frac{\dot{c}}{c}\right)\sigma_{xx}
\end{equation}

\noindent Now, we will see our shear satisfies the evolution equation
for the shear tensor:

\begin{equation}
\begin{split}
\frac{\partial\sigma_{xx}}{\partial t}&=\frac{\partial}{\partial
t}\left(\frac{\Sigma_{1}a^{2}}{S^{3}}\right)\\
&=\Sigma_{1}\left(\frac{2a\dot{a}}{S^{3}}\right)-3\frac{\Sigma_{1}a^{2}}{S^{4}}\dot{S}\\
&=\left(2\frac{\dot{a}}{a}-3\frac{\dot{S}}{S}\right)\sigma_{xx}\\
&=\left(\frac{\dot{a}}{a}-\frac{\dot{b}}{b}-\frac{\dot{c}}{c}\right)\sigma_{xx},
\end{split}
\end{equation}

\noindent We check it satisfies the evolution equation \eqref{shearevolution}.\\

\noindent Now, we present the generalized Friedmann equation. It is an equation that allows us to integrate the scale factor $S(t)$. We assume a $\gamma$-law for state equation $(p=(\gamma-1)\mu)$,

\begin{equation}
\mu=\frac{M}{S^{3\gamma}},
\end{equation}

\noindent where $\mu$ is the energy density and $\dot{M}=0$, the dot the covariant time derivative defined in \eqref{covariant time derivative}. So, it can be shown \cite{9812046} that the generalized Friedmann equation is given by:

\begin{equation}\label{generalizedfriedmann}
3\frac{\dot{S}^2}{S^2}=\frac{\Gamma^{2}}{S^{6}}+\frac{M}{S^{3\gamma}},
\end{equation}

\noindent where $2\Gamma^{2}=\Gamma_{1}^{2}+\Gamma_{2}^{2}+\Gamma_{3}^{2}$. When $\Gamma=0$ we get the usual Friedmann equation. The additional term at right is the shear Energy.\\

\noindent For late times the shear constant $\Gamma$ does not play a significant role in the evolution of scale factors, but at early times it has a great difference with the FLRW model. It can play an important role in physics processes in the early universe, such as Nucleosynthesis and structure formation \cite{wainwright}, \cite{thorne1967}, \cite{hawkingtayler}, \cite{olson}. For dust, we get an analytic solution for $S(t)$:

\begin{equation}
S(t)=\sqrt[3]{\frac{3}{4}Mt^{2}+\sqrt{3}\Sigma t}.
\end{equation}

\noindent It is a different expression that the one shown in \cite{9812046}, it can be checked it satisfies the Friedmann equation \eqref{generalizedfriedmann}. Using this expression for $S(t)$ we get:

\begin{equation}
W(t)=\frac{1}{\sqrt{3}\Sigma}\ln\left(\frac{t}{t+\frac{4\Sigma}{\sqrt{3}M}}\right),
\end{equation}

\noindent Thus, with these analytic solutions it is straightforward to obtain $a(t), b(t)$ and $c(t)$, and therefore the components of the shear tensor. We consider $M=1$ and three cases for the constants $\Sigma_1$, $\Sigma_{2}$ and $\Sigma_{3}$:

\noindent
\begin{enumerate}
\item $\Sigma_{1}=0$, $\Sigma_{2}=-\Sigma_{3}=0.05$.
\item $\Sigma_{1}=\Sigma_{2}=-0.05$, $\Sigma_{3}=0.1$.
\item $\Sigma_{1}=\Sigma_{2}=0.05$, $\Sigma_{3}=-0.1$.
\end{enumerate}

When it is considered the limit $t\rightarrow 0$, $S(t)\rightarrow 0$, there are two types of singularities, the \textit{cigar} singularity, which is the case 1 and 3 and the \textit{pancake} singularity which is the second case \cite{wainwright}, \cite{thorne1967}. The \textit{cigar} case means that two of the scale factors tend to $0$ while the third increases withouth bound, while the \textit{pancake} case means that one of the scale factors tend to $0$ and the other two increase.\\

\noindent In figures 1, 2 and 3 we present the evolution of Shear components for these three cases.
\FIGURE{\epsfig{file=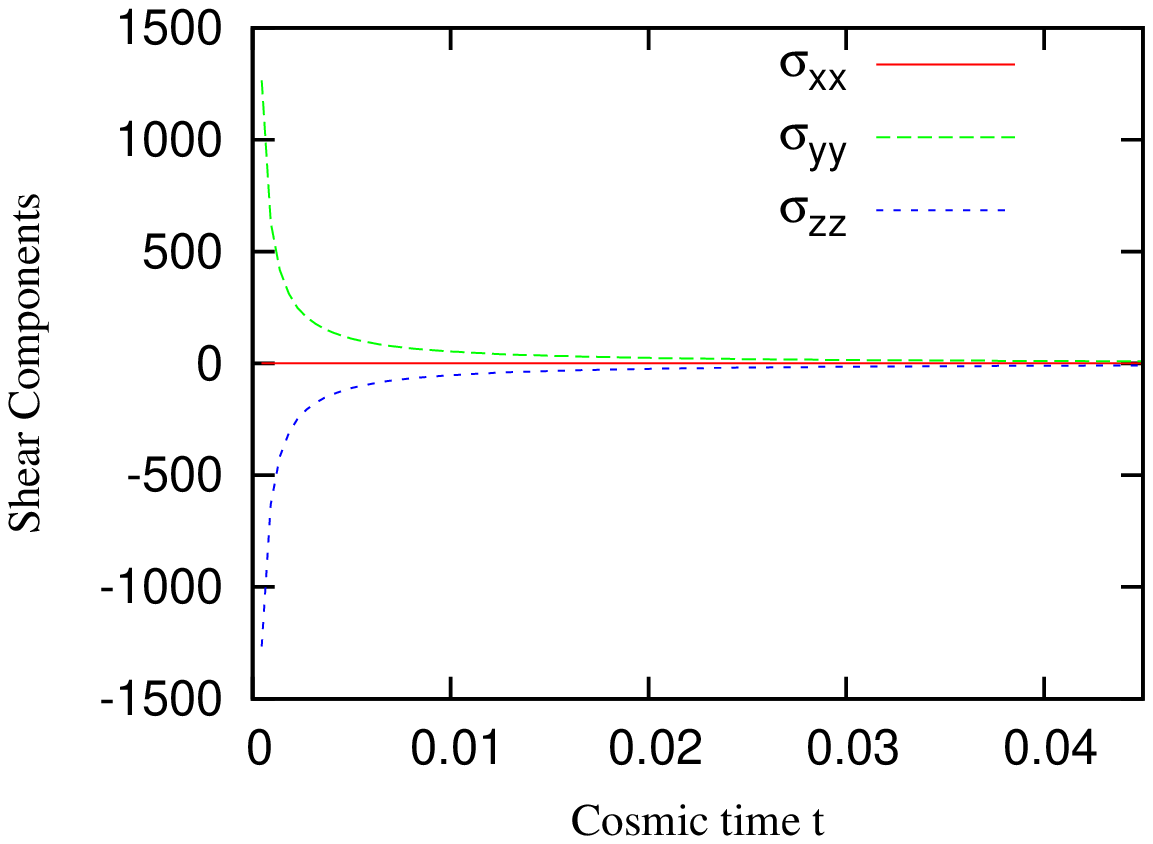,angle=-90,scale=0.7}
\caption[Example of figure]{Shear Evolution for case $\Sigma_{1}=0$, $\Sigma_{2}=-\Sigma_{3}=0.05$.}
\label{figura1}}
\smallskip
\FIGURE{\epsfig{file=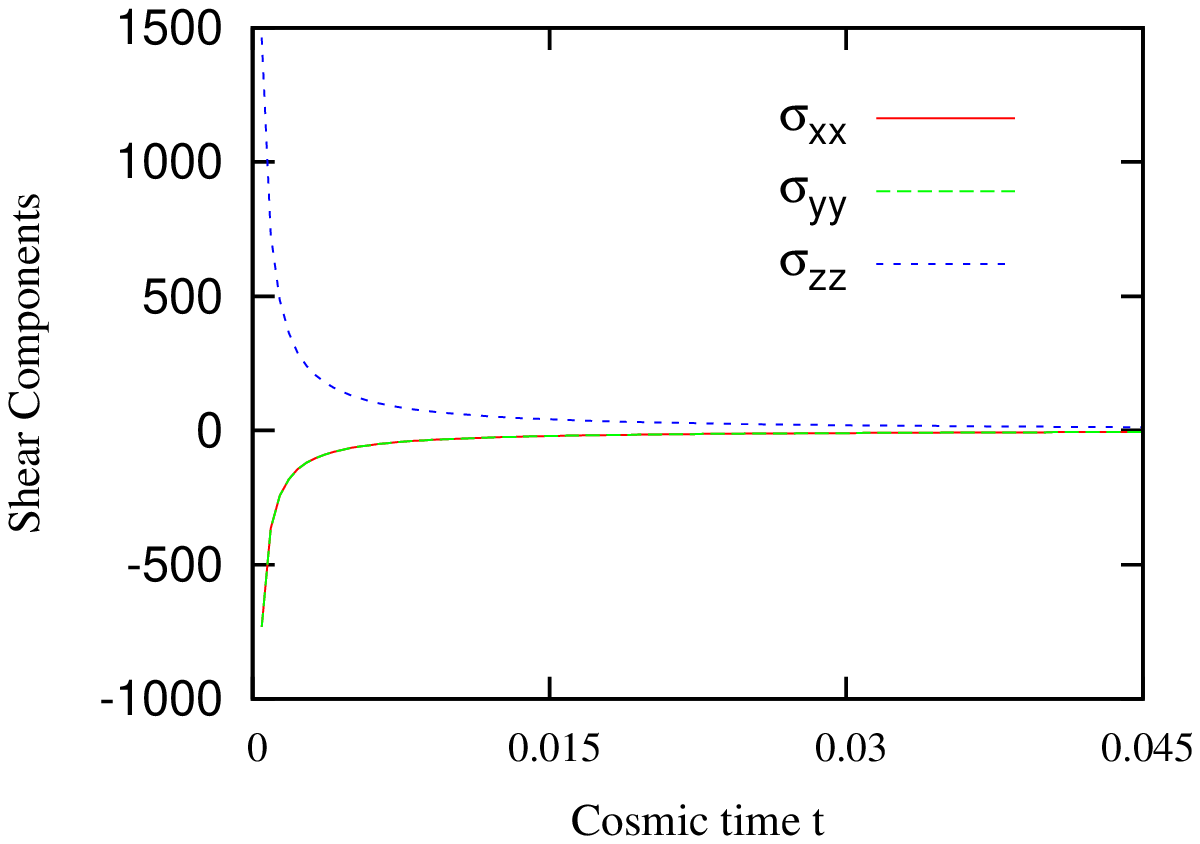,angle=-90,scale=0.7}
\caption[Example of figure]{Shear Evolution for case $\Sigma_{1}=\Sigma_{2}=-0.05$, $\Sigma_{3}=0.1$.}
\label{figura2}}
\smallskip
\FIGURE{\epsfig{file=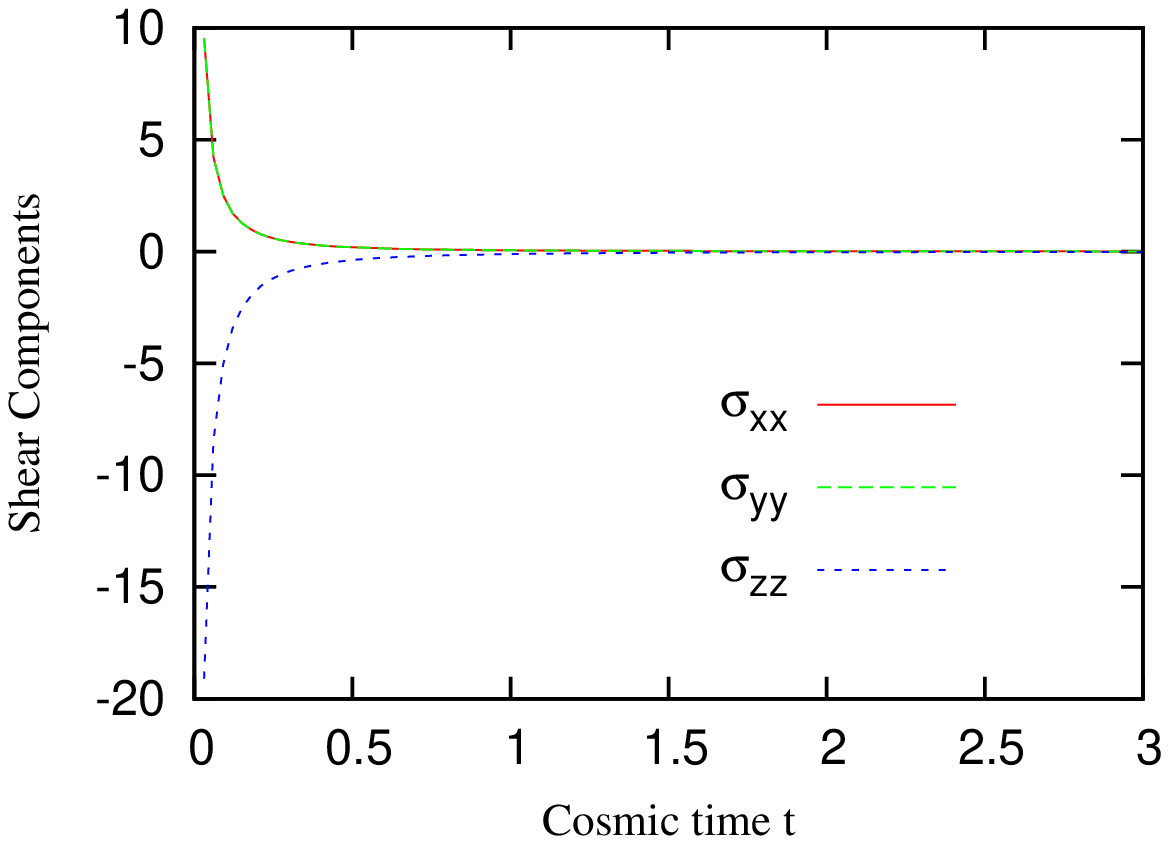,angle=-90,scale=0.7}
\caption[Example of figure]{Shear Evolution for case $\Sigma_{1}=\Sigma_{2}=0.05$, $\Sigma_{3}=-0.1$.}
\label{figura3}}

\FIGURE{\epsfig{file=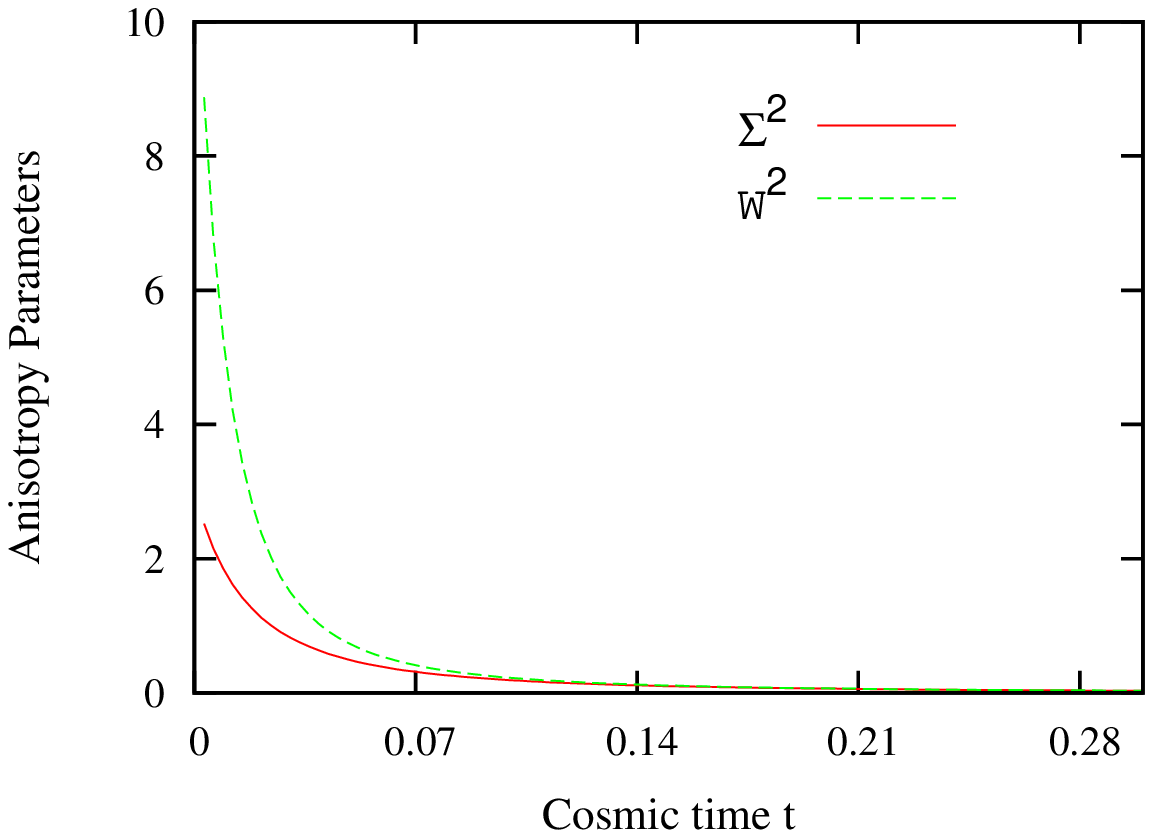,angle=-90,scale=0.7}
\caption[Example of figure]{Anisotropy parameters $\Sigma^2$ and $\mathcal{W}^{2}$ for case $\Sigma_{1}=0$, $\Sigma_{2}=-\Sigma_{3}=0.05$}
\label{figura4}}
\smallskip
\FIGURE{\epsfig{file=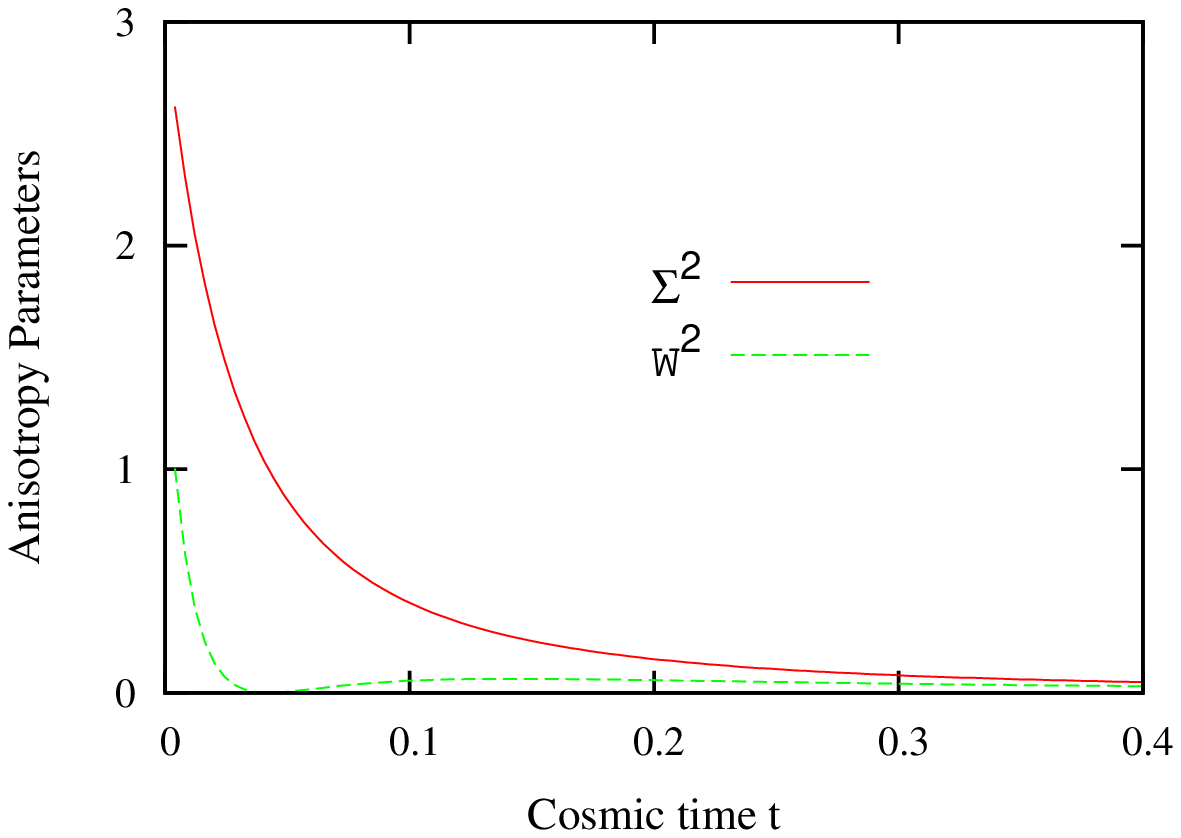,angle=-90,scale=0.7}
\caption[Example of figure]{Anisotropy parameters $\Sigma^2$ and $\mathcal{W}^{2}$ for case $\Sigma_{1}=\Sigma_{2}=-0.05$, $\Sigma_{3}=0.1$}
\label{figura5}}
\smallskip
\FIGURE{\epsfig{file=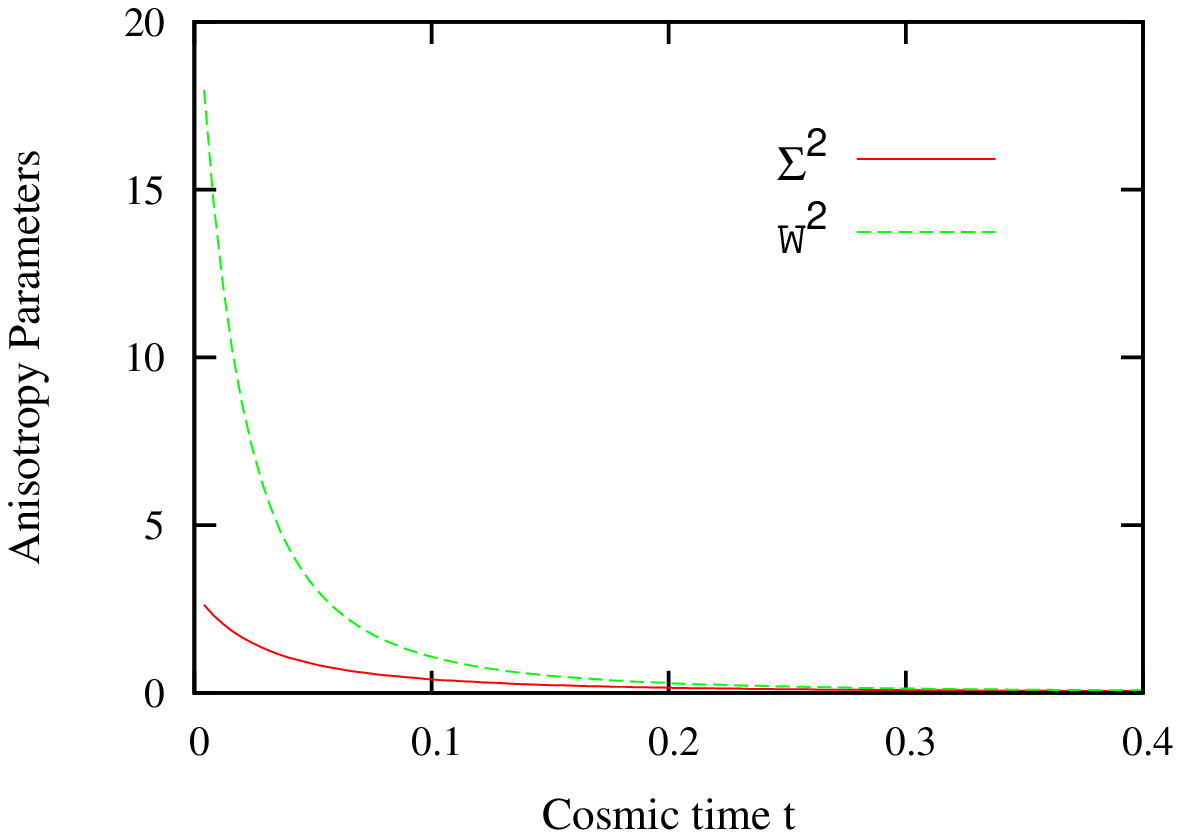,angle=-90,scale=0.7}
\caption[Example of figure]{Anisotropy parameters $\Sigma^2$ and $\mathcal{W}^{2}$ for case $\Sigma_{1}=\Sigma_{2}=0.05$, $\Sigma_{3}=-0.1$}
\label{figura6}}

\noindent We conclude these components tend to zero, as we could expect from the equation for shear tensor we got before. However, the appropiate way to know if this model isotropize is to consider the evolution of two scalars $\Sigma^{2}$ and
$\mathcal{W}^{2}$ \cite{9904252} defined as:

\begin{align}
&\Sigma^{2}=\frac{\sigma_{\alpha\beta}\sigma^{\alpha\beta}}{6H^{2}},\\
&\mathcal{W}^{2}=\frac{E_{\alpha\beta}E^{\alpha\beta}+H_{\alpha\beta}H^{\alpha\beta}}{6H^{4}}
\end{align}

\noindent These quantities are defined because the components of shear tensor are not dimensionless, they are normalized with the Hubble scalar $H$, and hence measuring the dynamical importance of the different variables with respect to the overall expansion of the universe.\\

\noindent When both parameters tend to zero we can say the model tends to isotropy. As was pointed out by \cite{9904252}, it was thought Bianchi $VII_{0}$ non-tilted dust model isotropize in the sense $\Sigma\rightarrow 0$ when $t\rightarrow\infty$. However, for $t\rightarrow\infty$ $\mathcal{W}\rightarrow\mathcal{W}_{0}$, where $\mathcal{W}_{0}$ is a constant whose value can be any positive number depending on the initial conditions \cite{9904252}. On the contrary these two factors tend to zero for our three cases of Bianchi I model. We illustrate these behaviours in figures 4, 5 and 6.\\

\noindent We see these parameters tend to zero in the three cases we have considered, the model isotropize. In the first case, as the constant $\Sigma_{1}=0$, the shear component $\sigma_{xx}=0$. However, the electric component $E_{xx}$, which we have not plotted here but we showed in \cite{caceres}, is not identically zero, although it is very small, compared with the other Electric components. Both Electric and Shear tensors are diagonal. In the other cases, where there exists axial simmetry, it is reflected in the components of shear tensor.

\section{Conclusions}
We have shown the shear tensor in BI cosmology and we analized the solutions in the dust model. From our analysis it is clear is convenient to be careful when the covariant derivative is considered.\\

\noindent The anisotropy parameters show us that this model isotropize, for different cases. For late times $S(t)\alpha\ t^{2/3}$, $H\alpha\ t^{-1}$, $\sigma_{\alpha\beta}\sigma^{\alpha\beta}\alpha\ S(t)^{-6}\alpha\ t^{-4}$, so $\Sigma^{2}\alpha\ t^{-2}$, while $E_{\alpha\beta}E^{\alpha\beta}\alpha\ S(t)^{-12}$, so $\mathcal{W}^{2}\alpha\ t^{-8}$ and as we see in the plots, these parameters decay, given a well defined behavior of the kinematical quantities in BI cosmology.

\end{document}